# Cooling Rate Effects in Sodium Silicate Glasses: Bridging the Gap between Molecular Dynamics Simulations and Experiments

Xin Li[1], Weiying Song[1], Kai Yang[1], N M Anoop Krishnan[1], Bu Wang[1], Morten M. Smedskjaer[2], John C. Mauro[3], Gaurav Sant[4,5,6], Magdalena Balonis[5,6], Mathieu Bauchy[1*]

[1]Physics of AmoRphous and Inorganic Solids Laboratory (PARISlab), University of California, Los Angeles, CA 90095-1593, U.S.A.

[2]Department of Chemistry and Bioscience, Aalborg University, 9220 Aalborg, Denmark

[3]Science and Technology Division, Corning Incorporated, Corning, NY 14831, U.S.A.

[4]Laboratory for the Chemistry of Construction Materials, University of California, Los Angeles, CA 90095-1593, U.S.A.

[5]Department of Materials Science and Engineering, University of California, Los Angeles, CA 90095-1593, U.S.A.

[6]Institute for Technology Advancement, University of California, Los Angeles, CA 90095-1593, California

[*]Corresponding author: Prof. Mathieu Bauchy, bauchy@ucla.edu

## Abstract

Although molecular dynamics (MD) simulations are commonly used to predict the structure and properties of glasses, they are intrinsically limited to short time scales, necessitating the use of fast cooling rates. It is therefore challenging to compare results from MD simulations to experimental results for glasses cooled on typical laboratory time scales. Based on MD simulations of a sodium silicate glass with varying cooling rate (from 0.01 to 100 K/ps), here we show that thermal history primarily affects the medium-range order structure, while the short-range order is largely unaffected over the range of cooling rates simulated. This results in a decoupling between the enthalpy and volume relaxation functions, where the enthalpy quickly plateaus as the cooling rate decreases, whereas density exhibits a slower relaxation. Finally, we demonstrate that the outcomes of MD simulations can be meaningfully compared to experimental values if properly extrapolated to slower cooling rates.

## I. Introduction

Molecular dynamics simulations are commonly used to investigate the structure and dynamics of disordered atomic networks[1–4], revealing information that often remains "invisible" to traditional experiments, despite recent advances in characterization techniques[5]. The structural characterization of disordered networks is complicated by the fact that glasses are out-of-equilibrium and continually relax toward the metastable supercooled liquid state[6]. As such, the structure and properties of glasses depend on their thermal history, e.g., the cooling rate through the glass transition regime[4]. This raises serious questions regarding the reliability of MD simulations, since the cooling rate used in such simulations is typically on the order of $10^{12}$ K/s, that is, much higher than that typically achieved experimentally (1 to 100 K/s).

Several numerical studies have investigated the effect of cooling rate on the structure and properties of model glasses[7] and silica[4,8,9], among others. In MD simulations of silica, it is typically observed that the usage of high cooling rates results in the formation of unrealistic structural defects[9], although the overall structure only weakly depends on the thermal history[4]. In contrast, some properties, e.g., the density and thermal expansion coefficient, were found to be highly sensitive to the cooling rate[8]. However, silica features various unique properties, including (i) a high connectivity, which inhibits structural relaxation at low temperature[10], (ii) a low fragility index[11], which limits the propensity for relaxation within the supercooled liquid state close to the glass transition temperature ($T_g$), and (iii) an anomalous volume-temperature (*V*–*T*) diagram at high temperature[6]. Since all these features can affect the relaxation of the system upon cooling, it remains unclear whether the cooling-rate-dependence of the structure and properties of silica is representative of that of more depolymerized and fragile silicate glasses. More generally, this raises the following questions: Are the glasses prepared by MD simulations representative of real glasses formed in experiment? Can their outcomes be reliably extrapolated to the low cooling rates typically used experimentally?

To address these questions, here we perform MD simulations of a sodium silicate glass, a archetypical model for various multi-component silicate glasses of industrial[12–15] and geological[16] relevance, prepared over a range of cooling rates accessible to MD, ranging from 100 to 0.01 K/ps. We show that although the short-range order structure remains largely unaffected by the cooling rate, medium-range structural features like the $Q^n$ distribution exhibit a greater dependence on the cooling rate. These distinct behaviors result in a decoupling between the relaxation of enthalpy and volume with decreasing cooling rate. Although MD simulations are limited to high cooling rates, they offer predictions that can be quantitatively compared to experimental values, provided that they are properly extrapolated to lower cooling rates.

# II. Simulations details

## 1. Preparation of the glasses

To establish our conclusions, we perform MD simulations of a sodium silicate glass $(Na_2O)_{30}(SiO_2)_{70}$ (mol %) comprising 3000 atoms. The system is simulated via the classical MD approach, using the well-established Teter potential[17–19]. Following previous studies, a short-range repulsive term is added to avoid the “Buckingham catastrophe” at high temperature[1,19]. Cutoffs of 8 and 12 Å are used for the short-range and Coulombic interactions, respectively. The Coulombic interactions are evaluated with the PPPM algorithm with an accuracy of $10^{-5}$. All simulations are carried out using the LAMMPS package[20] with a fixed time step of 1.0 fs.

The glasses are simulated using the traditional melt-quench procedure as follows. First, the atoms are randomly placed within a cubic box while ensuring the absence of any unrealistic overlap. The system is then melted at 4000 K in the canonical (*NVT*) ensemble for 10 ps and at zero pressure (*NPT* ensemble) for 100 ps, which ensures complete loss of the memory of the initial configuration. The system is subsequently cooled linearly to 300 K at zero pressure (*NPT* ensemble) with five cooling rates spanning four orders of magnitude: 100 K/ps, 10 K/ps, 1 K/ps, 0.1 K/ps, and 0.01 K/ps. This represent a significant computational effort, as most MD simulations typically rely on cooling rates equal to or larger than 1 K/ps. All of the resulting glasses are further relaxed at 300 K and zero pressure for 100 ps before an *NVT* run of 100 ps for statistical averaging. In the following, all properties referring to the "glassy state" are averaged over 100 configurations extracted from this run and error bars are evaluated based on the standard deviations.

### 2. Glass transition and inherent configurations

To determine where the system falls out of equilibrium, i.e., the onset of the glass transition, we compute the enthalpy of the inherent configurations (noted ground-state enthalpy in the following) explored by the system as a function of temperature. This is achieved as follows. First, for each cooling rate, 16 independent configurations are *a posteriori* extracted during the cooling process at every 100 K interval. Each configuration is then subjected to an energy minimization. The ground state enthalpy is subsequently calculated by averaging the final enthalpies of the 16 relaxed configurations. Finally, for each cooling rate, the fictive temperature ($T_f$, see Sec. III.1) is determined by the linearly fitting the low- and high-temperature domains of the ground-state enthalpy and determining the temperature at which these lines intercept.

### 3. Structural characterization

#### a. Pair distribution functions

To compare the structure of the simulated glass to experimental data from neutron diffractions and assess the impact of cooling rate on the glass structure, we first compute each partial pair distribution functions $g_{ij}(r)$. The neutron pair distribution function is then calculated as:

$$g_{\mathrm{N}}(r) = (\sum_{i,j=1}^{n} c_i c_j b_i b_j)^{-1} \sum_{i,j=1}^{n} c_i c_j b_i b_j g_{ij}(r) \quad \text{Eq. 1}$$

where $c_i$ is the fraction of $i$ atoms ($i$ = Na, Si, or O) and $b_i$ the neutron scattering length of the species (given by 3.63, 4.1491, and 5.803 fm for Na, Si, and O atoms, respectively). To enable a meaningful comparison with experimental data[21], the neutron pair distribution function is subsequently broadened to account for the maximum of wave vector ($Q_{\max}$ = 22.88 Å) used experimentally[21]. This is achieved by convoluting the computed neutron pair distribution function with a normalized Gaussian distribution with a full width at half-maximum (FWHM) given by FWHM = 5.437/$Q_{max}$[21].

### b. Structure factors

To investigate the structure of the glass over intermediate length scales, we compute the partial structure factors $S_{ij}(Q)$ from the Fourier transform of the partial pair distribution functions:

$$S_{ij}(Q) = 1 + \rho_0 \int_0^R 4\pi r^2 \left(g_{ij}(r) - 1\right) \frac{\sin(Qr)}{Qr} F_L(r) dr, \quad \text{Eq. 2}$$

where $Q$ is the scattering vector, $\rho_0$ is the average atomic number density, and $R$ is half of the simulation box length. The $F_L(r) = \sin(\pi r/R)/(\pi r/R)$ term is a Lortch-type window function used to reduce the effect of the finite cutoff of *r* in the integration [19]. The use of this function reduces the ripples at low Q but induces a broadening of the structure factor peaks. The total neutron structure factor is then evaluated from the partial structure factors following:

$$S_N(Q) = \left(\sum_{i,j=1}^{n} c_i c_j b_i b_j\right)^{-1} \sum_{i,j=1}^{n} c_i c_j b_i b_j S_{ij}(Q) \quad \text{Eq. 3}$$

### c. Coordination numbers

The connectivity of the network is assessed by computing the coordination numbers of Si, O, and Na atoms. This is achieved by enumerating for each atom the number of neighbors within its first coordination shell, with a cutoff chosen as the first minimum after the first peak of the partial pair distribution function. This analysis is used to discriminate bridging oxygen (BO) and non-bridging oxygen (NBO) atoms, which exhibit either two or one Si atoms in their first coordination shell, respectively.

## 4. Accuracy of the potential

The ability of the Teter potential to predict the features of sodium silicate glasses has already extensively been discussed[17–19,22,23]. In particular, it has been shown to yield realistic structural[17,24], dynamical[16,25,26], vibrational[19], thermodynamical[19,27,28], and mechanical[29–32] properties. To illustrate these features, Fig. 1 shows the broadened neutron pair distribution function and structure factor[19] computed for the glass prepared with a cooling rate of 0.01 K/ps, which are compared to neutron diffraction data[33]. As shown in Fig. 1, both the neutron pair distribution function and structure factor show a very good agreement with the experimental data.

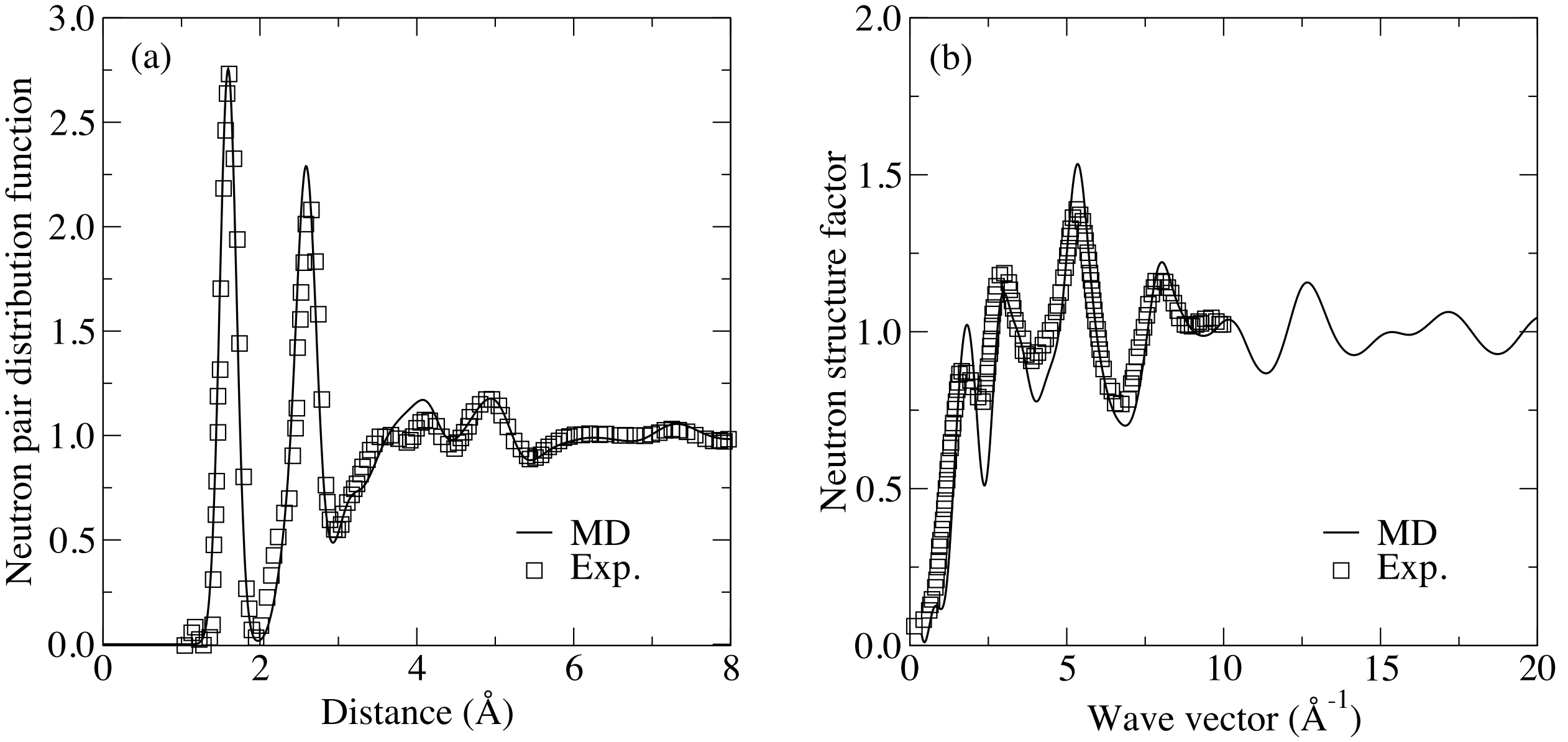

*Fig. 1: Computed (a) neutron pair distribution function and (b) structure factor of a $(Na_2O)_{30}(SiO_2)_{70}$ glass prepared with a cooling rate of 0.01 K/ps. The results are compared to neutron diffraction data[33].*

# III. Results

## 1. Glass transition

We first focus on the temperature dependence of the ground-state enthalpy (see Sec. II.2) to ensure that our methodology yields realistic results regarding the influence of cooling rate. As shown in Fig. 2, we observe that the ground-state enthalpy decreases monotonically with decreasing temperature. At a given temperature, a change of slope is observed, indicating that the system falls out of equilibrium. This corresponds to the point at which the relaxation time of the system starts to exceed the observation time, that is, to the onset of the glass transition regime. In the following, we refer to this temperature as the "fictive temperature" of the glass, since the "glass transition temperature" ($T_g$) is typically defined as the temperature as which the viscosity of the glass reaches $10^{12}$ Pa·s[6], a quantity that cannot be directly assessed by conventional MD simulations. In contrast to glassy silica, the ground-state enthalpy of sodium silicate continues to decrease in the glassy state (low temperature), which indicates that, in agreement with recent experiments and simulations[34,35], some extent of structural relaxation continues to occur below the glass transition temperature.

The simulations reproduce typical features of the glass transition that are observed experimentally[6]. In particular, the use of a slower cooling rate results in (i) lower values of fictive temperatures (see Fig. 12 in Sec. IV.1) and (ii) lower values of the final ground-state enthalpy in the glassy state (see the inset of Fig. 2). As expected, this indicates that (i) as the cooling rate decreases (i.e., as the observation time increases), the system can remain in the metastable equilibrium supercooled liquid state until lower temperatures and (ii) the glass can achieve a more stable (lower ground-state enthalpy) configuration as the cooling rate decreases, since it has more time to sample the enthalpy landscape to find a lower enthalpy configuration[36]. In our simulations, the difference of ground-state enthalpies between the glasses obtained using the highest and lowest cooling rates is on the order of 10 kJ/mol (40 meV/atom).

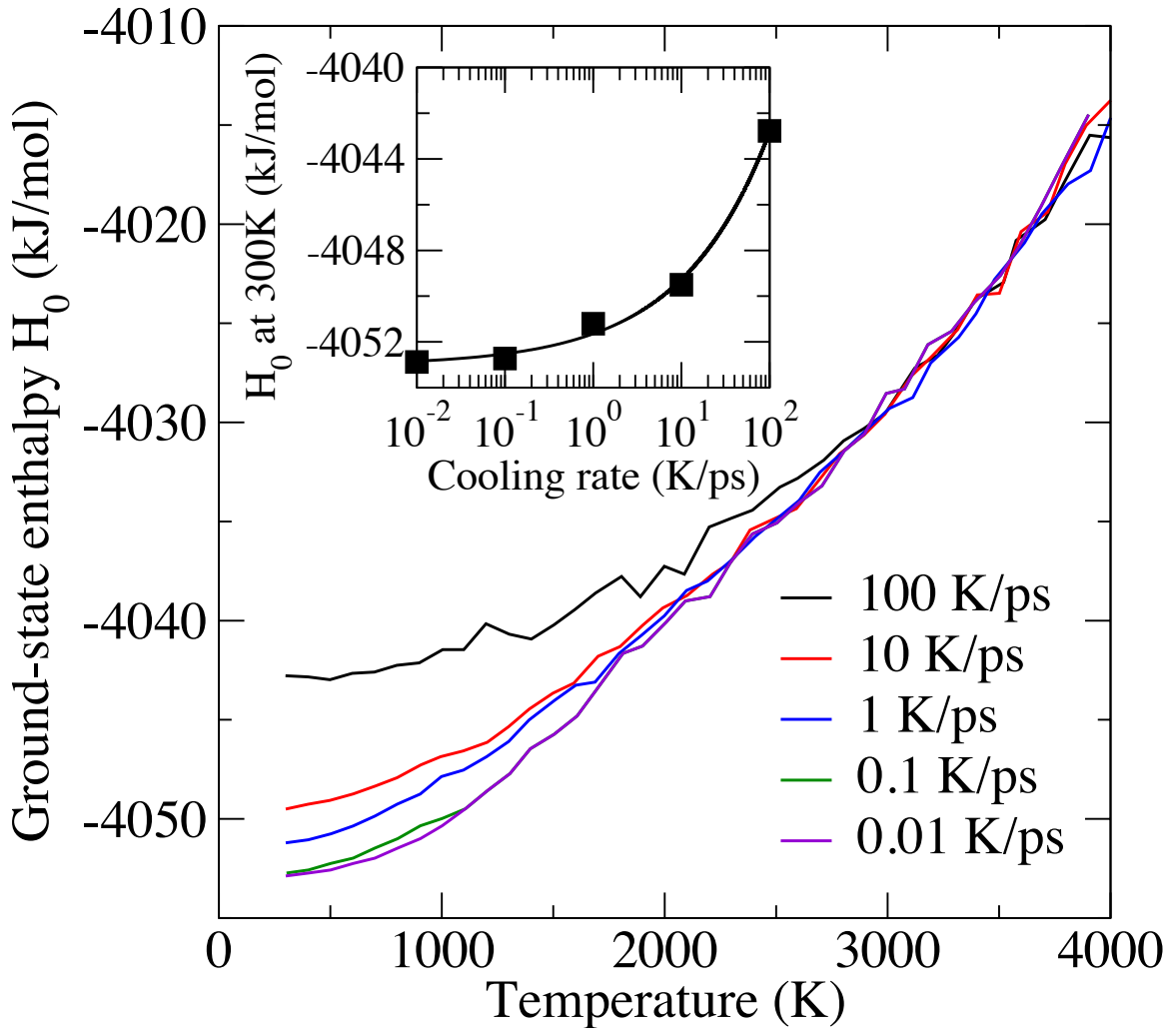

*Fig. 2: Ground-state enthalpy ($H_0$) as a function of temperature under different cooling rates. The inset shows $H_0$ in the glassy state as a function of the cooling rate. The line is a guide for the eye.*

As shown in Fig. 3, the glass transition also manifests itself by a change of slope within the temperature-dependence of the molar volume, although we note that the transition does not appear as sharp as in the case of the ground-state enthalpy. Nevertheless, our simulations reproduce the trend that is typically observed experimentally, i.e., the molar volume of the glass decreases with decreasing cooling rate (see the inset of Fig. 3). An opposite trend can be observed in glassy silica[4,8], depending on the range of cooling rates under study, i.e., the molar volume of the glass can increase with decreasing cooling rate. This is a consequence of the anomalous behavior of supercooled liquid silica, which features a minimum of molar volume (observed around 4800 K with the BKS potential[4,8]). In that respect, sodium silicate behaves in a more conventional way than silica and, as such, offers a more general archetypical model to assess the general role of the cooling rate in controlling glass properties.

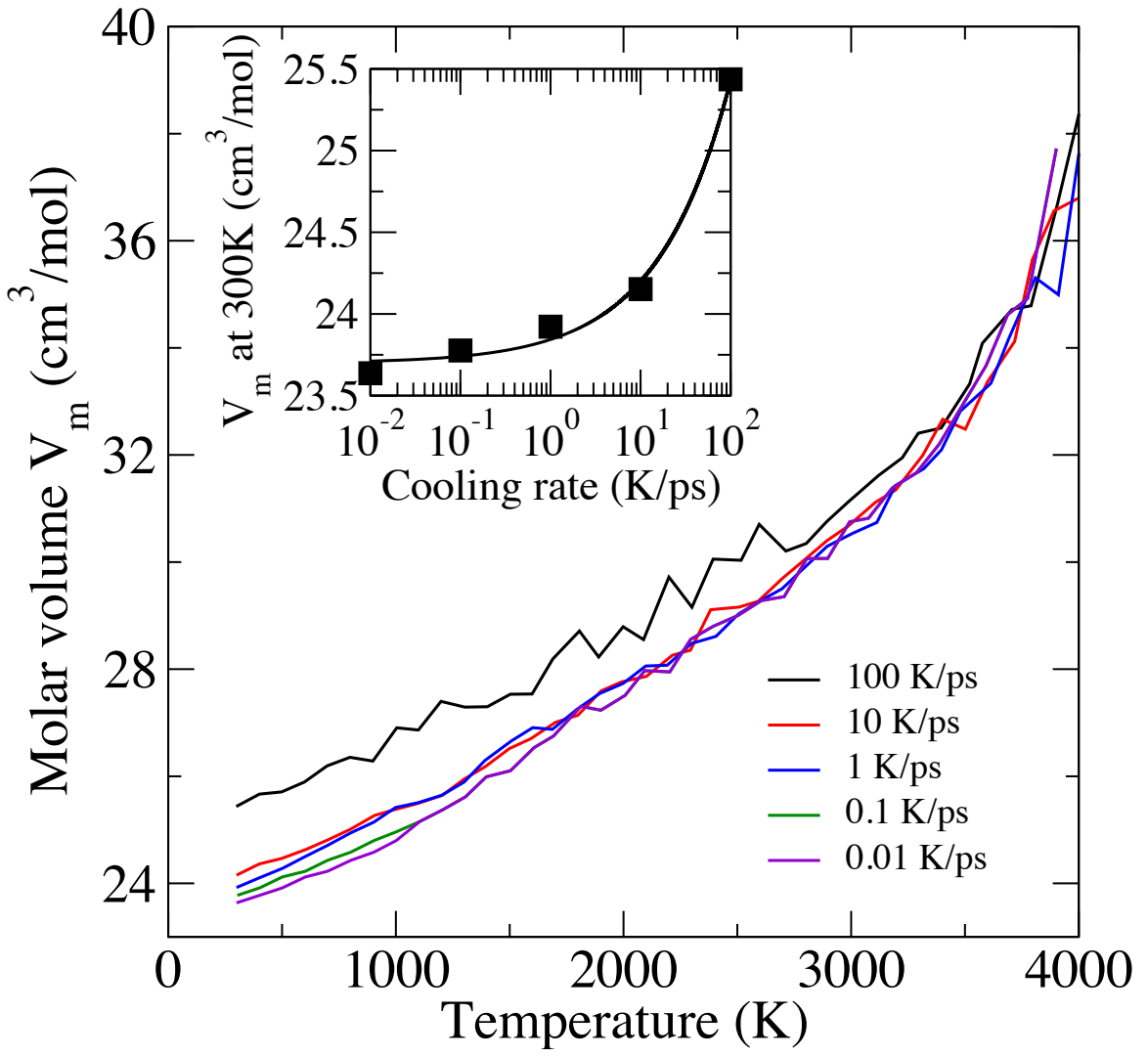

*Fig. 3: Molar volume ($V_m$) as a function of temperature under different cooling rates. The inset shows $V_m$ in the glassy state as a function of the cooling rate. The line is a guide for the eye.*

A linear fit of low temperature part of the molar volume curves from Fig. 3 yields the volumetric coefficient of thermal expansion (CTE)[37], which is shown in Fig. 4 as a function of the cooling rate. Our simulations overestimate the value of the CTE by about a factor of two[38,39], which has been noted to be a limitation of the Teter potential[37]. Overall, we do not observe significant relative variations in the CTE across the range of cooling rates simulated. As shown in Fig. 4, this behavior strongly differs from the trend observed for glassy silica[8], wherein the CTE drops toward zero as the cooling rate decreases. In contrast, the present data suggest that the CTE of sodium silicate only slightly decreases toward the experimental value at low cooling rate. This different behavior can be understood from the fact that, in contrast to silica, sodium silicate continues to relax even below the fictive temperature (see the continuous decrease of energy in Fig. 2). Such low temperature relaxation appears to reduce the influence of the cooling rate on the thermal expansion of the glass.

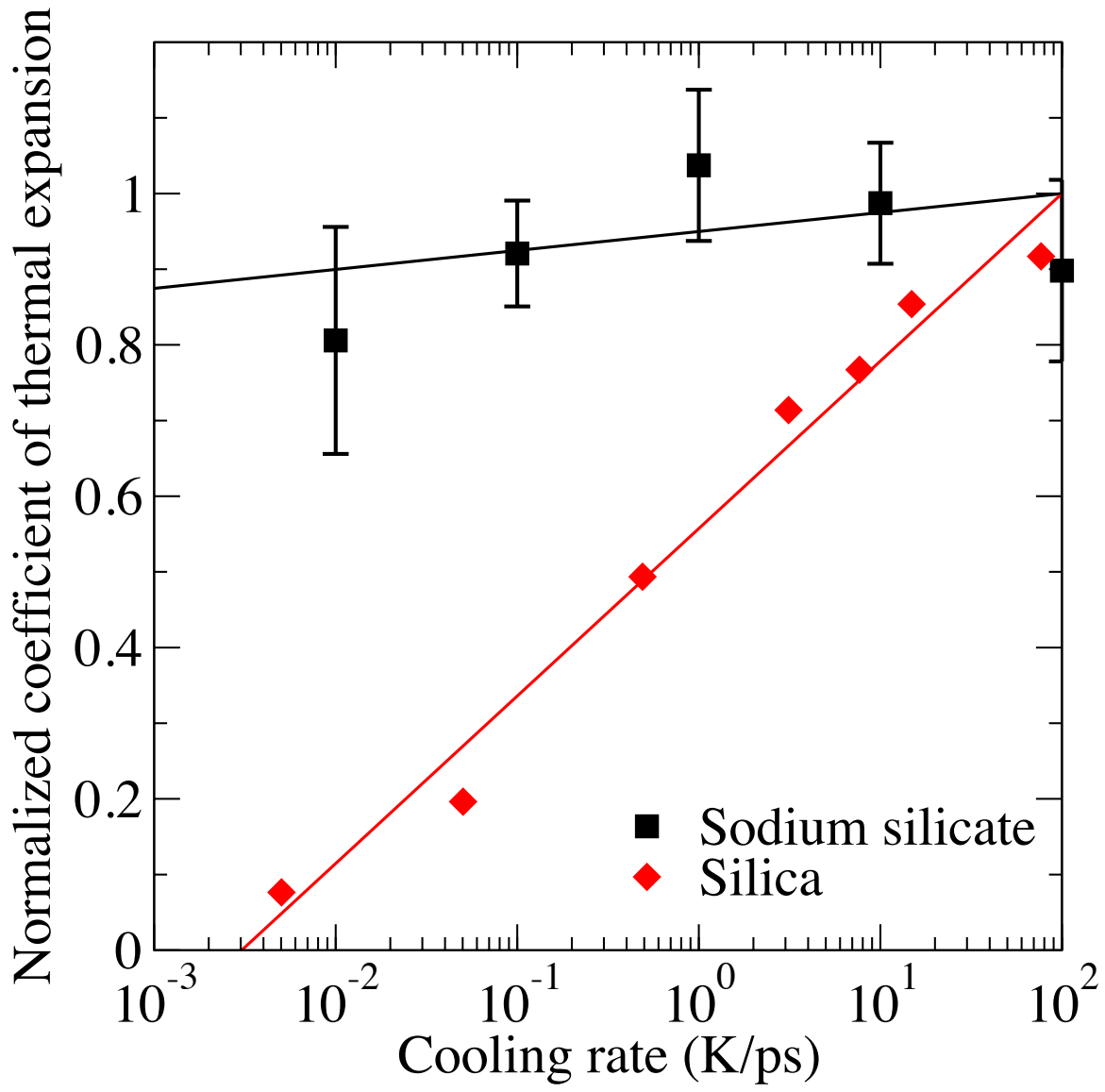

*Fig. 4: Computed coefficients of thermal expansion (CTE) of sodium silicate glasses (normalized by the CTE for 100 K/ps) as a function of the cooling rate. The data are compared to those obtained for glassy silica[8] (normalized by the CTE for 100 K/ps). The lines serve as guides for the eye.*

## 2. Short-range structural order

### a. Pair distribution function

We now direct our attention to the short-range order (< 3 Å) structure of the formed glasses. Fig. 5 shows the Si–O and Na–O partial pair distribution functions (PDFs). No significant shift is observed within the first and second coordination shell peaks. This suggests that the cooling rate does not significantly affect the local environment (bond distance, coordination number) of Si and Na cations. This contrasts with the results obtained for glassy silica, wherein the Si–O PDF features some significant variation with respect to the cooling rate[4]. On the other hand, as shown in Fig. 6, the Si–Si and Na–Na PDFs exhibit a more pronounced dependence on the cooling rate. For both PDFs, the first peak becomes sharper upon slower cooling. This denotes the formation of a more ordered atomic network. Also, the first peak of the Si–Si PDF shifts toward higher distance as the cooling rate decreases. Since this Si–O bond distance remains largely unaffected by the cooling rate (see  Fig. 5a), this suggests that the Si–O–Si bond angle is increasing for lower cooling rates (see Sec. III.2.b). In contrast, we observe that the first peak of the Na–Na PDF shifts toward lower distance as the cooling decreases. This suggests that Na atoms tend to “cluster” together inside pockets with lower cooling rates, in agreement with the picture of Greaves’ modified random network model[25,40].

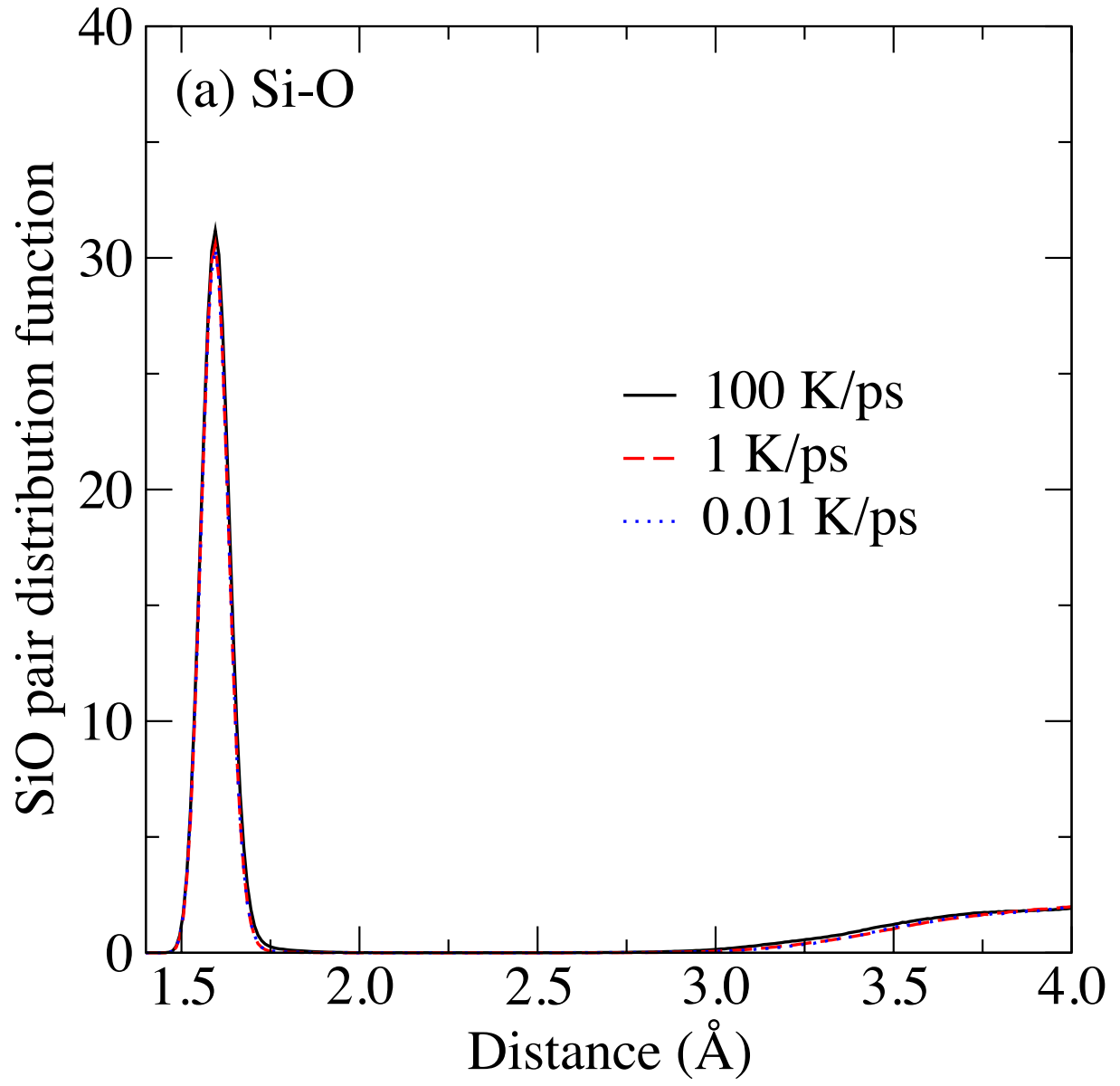

*Fig. 5: (a) Si–O and (b) Na–O partial pair distribution functions in the glassy state for selected cooling rates.*

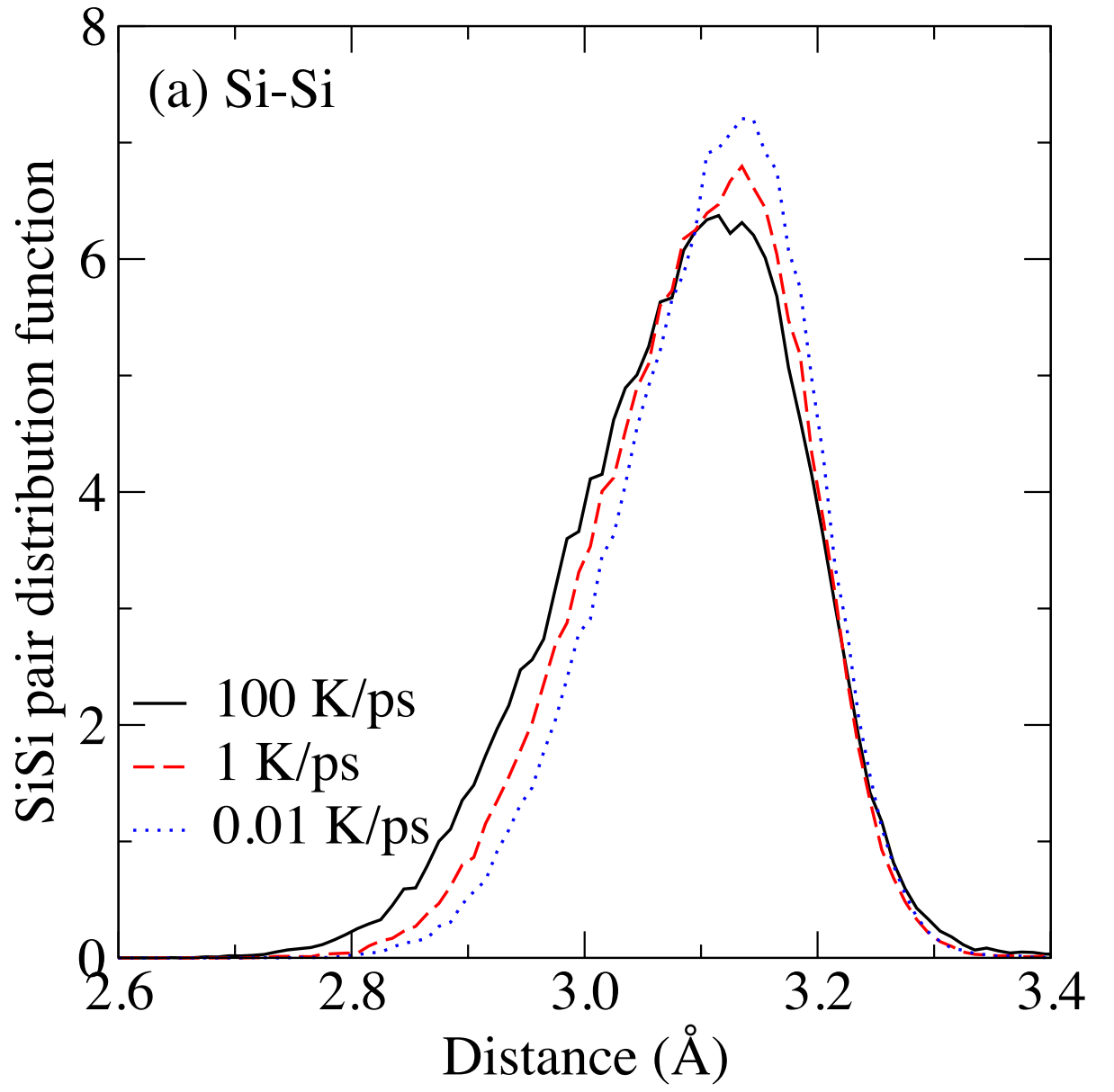

*Fig. 6: (a) Si–Si and (b) Na–Na partial pair distribution functions in the glassy state for selected cooling rates.*

### b. Angles

The fact that the Si–Si PDF exhibits some dependence on the cooling rate while the Si–O one remains largely unaffected suggests that thermal history might have a more significant influence on bond angles than on bond lengths. Fig. 7 shows the O–Si–O (intra-tetrahedral) and Si–O–Si (inter-tetrahedral) bond angle distributions (BADs). First, we observe that both BADs become significantly sharper as the cooling rate decreases (see the insets of Fig. 7), which suggests that lower cooling rates result in higher angular order. Second, we note that the average value of the O–Si–O angle remains largely unaffected by the cooling rate, with a value that is close to 109°, as expected from the tetrahedral geometry of the $SiO_4$ units. Similar conclusions are obtained in the case of glassy silica[8]. In contrast, we observe that the Si–O–Si BAD exhibits a significant shift toward higher angles as the cooling rate decreases. A similar behavior is observed in silica[4,8] and calcium aluminosilicate glasses[41,42], which suggests that this feature is general to silicate glasses. In the case of silica, the opening of the Si–O–Si angle indicates an opening of the network[4,8], in agreement with the decrease of the density as the cooling rate decreases. Here, this opening appears to be in contradiction with the observed increase of the density with lower cooling rates (see Fig. 3). This discrepancy might be explained by the fact that, in the case of sodium silicate, the opening of the Si–O–Si angle could facilitate the formation of large pockets of Na. As such, although the opening of the inter-tetrahedral angle mostly results in the creation of some empty space in silica, network modifiers species like Na cations can fill these voids, which ultimately results in a more efficient overall atomic packing.

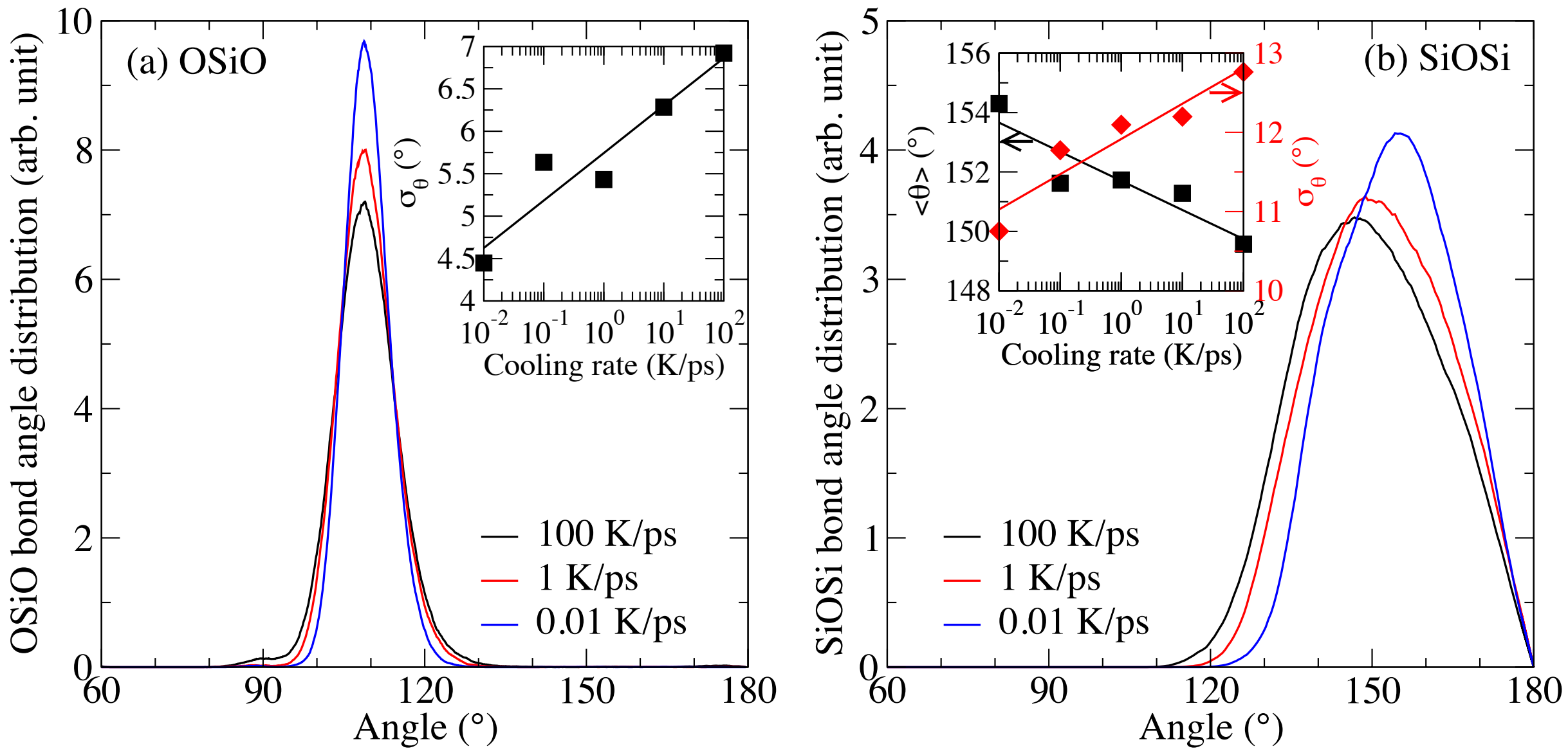

*Fig. 7: (a) Intra-tetrahedral O–Si–O bond angle distribution (BAD) in the glassy state for selected cooling rates. The inset shows the standard deviation $\sigma_\theta$ of the BADs as a function of the cooling rate. The line is a guide for the eye. (b) Inter-tetrahedral Si–O–Si BAD in the glassy state for selected cooling rates. The inset shows the average value (<θ>, left axis) and standard deviation value ($\sigma_\theta$, right axis) of the BADs as a function of the cooling rate. The lines serve as guides for the eye.*

### c. Connectivity

We now investigate the effect of the cooling rate on the connectivity of the glass network. As shown in Fig. 8, we observe the formation of a few (< 1%) over-coordinated Si species at high cooling rate. Similar observations are made in the case of glassy silica[4]. Such coordination defects are found mostly to disappear at lower cooling rates, although a small fraction of defects is still found. On the other hand, we observe that the coordination number of Na cations noticeably increases with lower cooling rates. This agrees with our observation that Na cations tend to form some clusters inside the large pockets created within the network following the opening of the Si–O–Si angle.

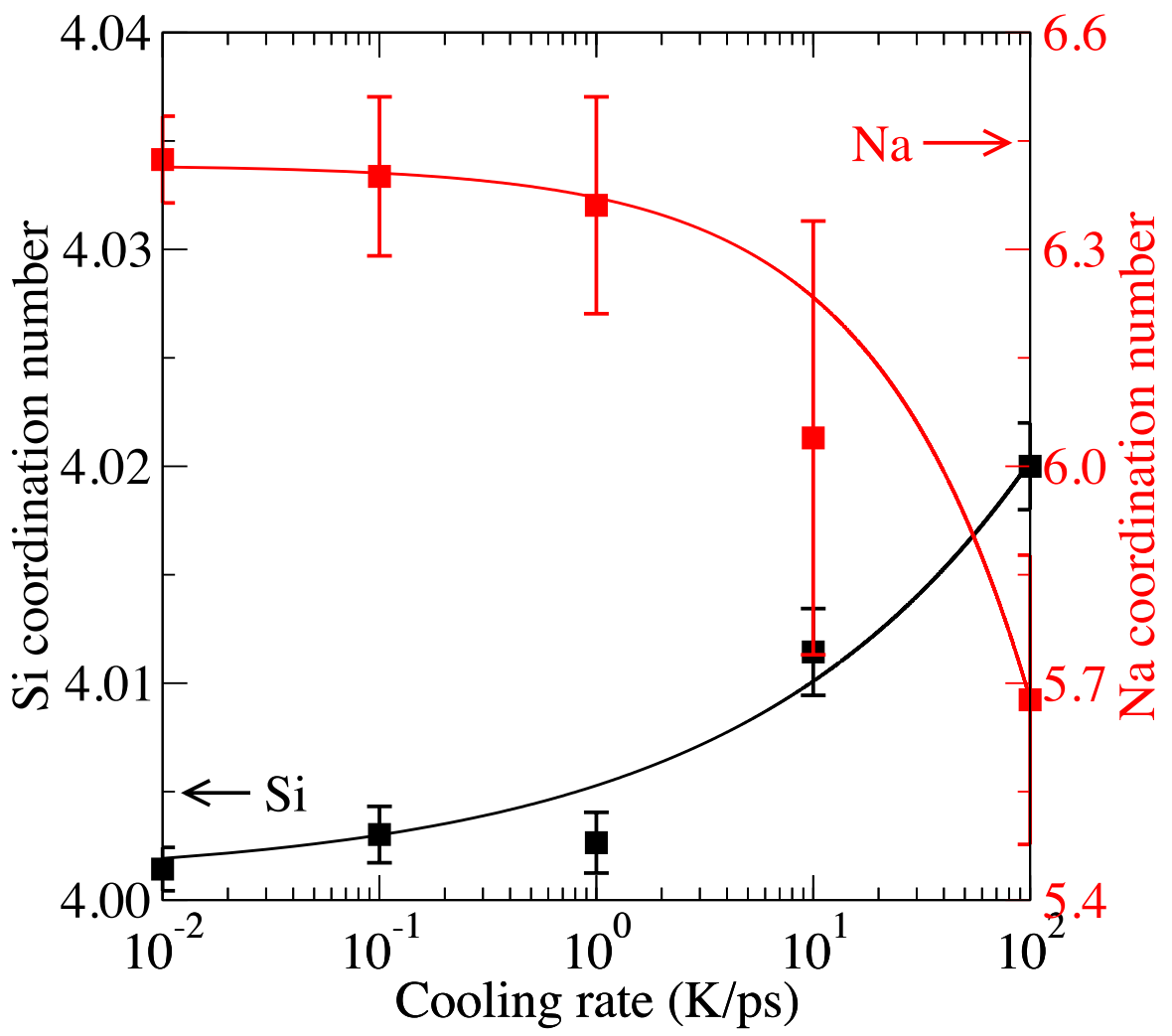

*Fig. 8: Average coordination number of Si (left axis) and Na (right axis) atoms in the glassy state as a function of the cooling rate. The lines serve as guides for the eye.*

We next focus on the O species. In glassy silica, the $SiO_4$ tetrahedra are connected to each other through their corners, so that the network only features bridging oxygen (BO) species[6]. In contrast, Na cations tend to depolymerize the network by breaking Si–BO–Si bonds. Namely, each added Na cation results in the creation of a non-bridging oxygen (NBO) species through the formation of Si–$O^-Na^+$ bonds[6]. As shown in Fig. 9, we observe that the fraction of NBO increases as the cooling rate decreases and quickly plateaus toward the theoretical value at low cooling rate, i.e., assuming that the number of NBOs equals that of Na cations.

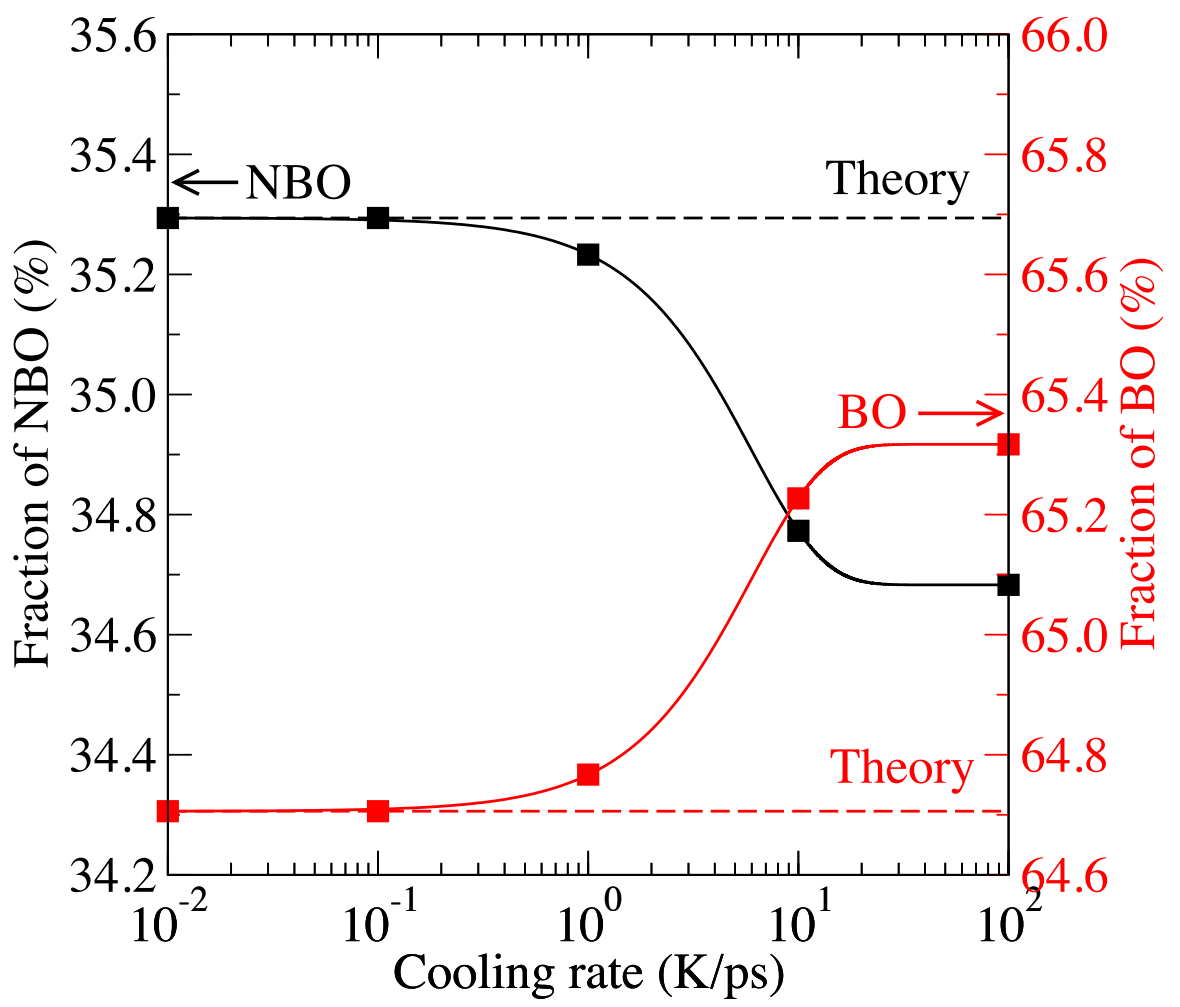

*Fig. 9: Fraction of non-bridging (NBO, left axis) and bridging oxygen (BO, right axis) in the glassy state as a function of the cooling rate. The solid lines serve as guides for the eye. The dashed lines indicate the theoretical values, assuming that each Na atom creates 1 NBO.*

## 3. Medium-range structural order

We now turn our attention to the medium-range order structure (3 < $r$ < 10 Å) of the sodium silicate glasses. We first focus on the connectivity around $SiO_4$ tetrahedra units, which can be assessed from the knowledge of the $Q^n$ distribution, where a $Q^n$ species denotes a $SiO_4$ tetrahedral unit connected to $n$ other Si atoms, i.e., comprising $n$ BO atoms. Fig. 10a shows the fraction of the $Q^n$ species as a function of the cooling rate. These data are compared to the outcomes of a fully random model (wherein the NBO are randomly attributed to each Si atoms, see Ref.[19] for details) and a fully ordered model (wherein each Si atom does not exhibit more than one NBO), as shown in Fig. Fig. 10b. We observe that, at high cooling rates, the distribution of the $Q^n$ units is in close agreement with the prediction from the random model. However, these results largely differ from experimental data[43] (see Fig. 13), which present an excess of $Q^3$ units with respect to the predictions of the random model[19]. Interestingly, we observe that the fraction of $Q^3$ units increases toward the prediction from the ordered model with decreasing cooling rate, although it remains low compared to the experimental value (around 68%[43], see Fig. 13). This suggests that as the cooling rate decreases, the atomic network self-organizes to favor the formation of $Q^3$ units through the transformation $Q^2 + Q^4 \rightarrow 2Q^3$ (see Sec. IV.1 for a discussion on this point).

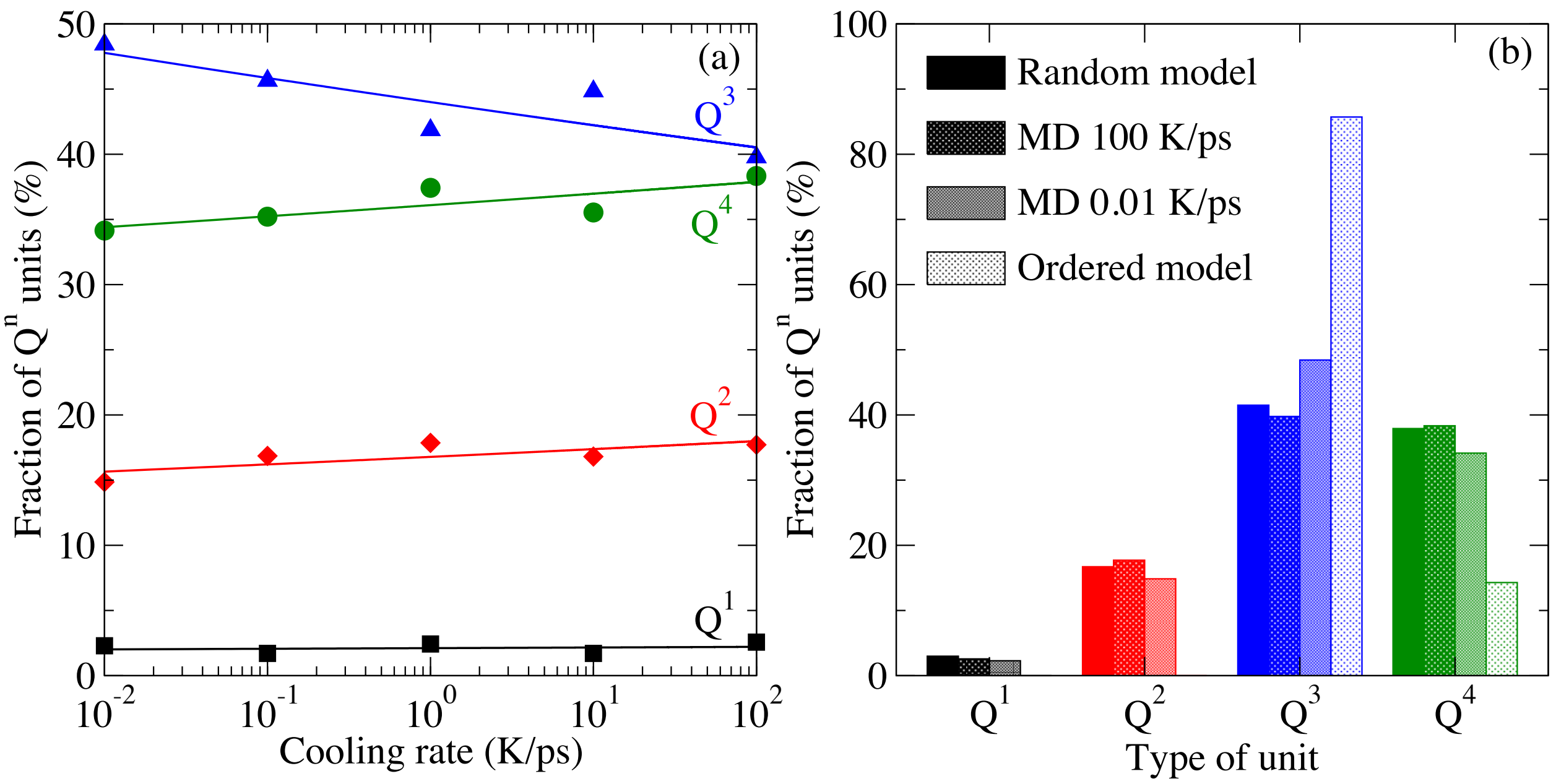

*Fig. 10: (a) Fraction of the $Q^n$ species in the glassy state as a function of the cooling rate. The lines serve as guides for the eye. (b) Comparison of the computed data (for a cooling rate of 100 K/ps and 0.01 K/ps) with those obtained from fully random and fully ordered models [19].*

Although they contain the same information as the partial PDFs, the partial structure factors place higher emphasis on structural correlations at longer distances and, hence, can also be used to assess the effect of the cooling rate on the medium-range order. As shown in Fig. 11, we observe that the partial structure factors are more affected by the cooling rate than the PDFs. In particular, the first-sharp diffraction peak (FSDP), which captures structural correlations within the medium-range order[24,44], exhibits significant variations in both its height and position with varying cooling rates. The FSDP of both the Si–Si and O–O partial structure factors is found to sharpen at lower cooling rate, which denotes an increased degree of order within the medium-range order of the atomic network. In both cases, we also observe a shift of the FSDP toward lower values of reciprocal wave vector, as observed in glassy silica[4]. This is in agreement with the opening of the silicate network resulting from the increase of the average inter-tetrahedral angle.

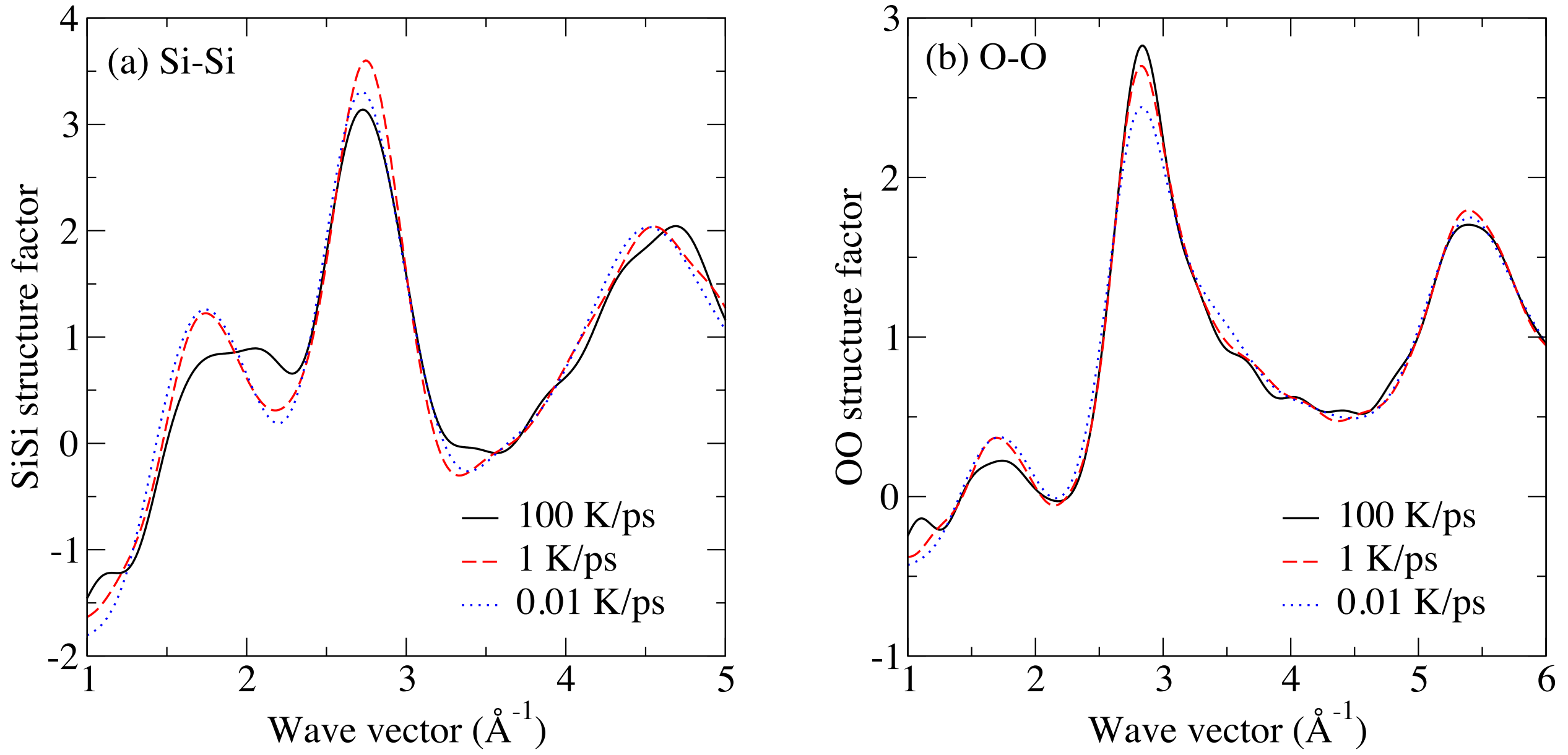

*Fig. 11: (a) Si–Si and (b) O–O partial structure factors in the glassy state for selected cooling rates.*

# IV. Discussion

## 1. Linking MD simulations and experiments

We now discuss how the predictions of the present MD simulations can be compared to experimental data despite the vast difference in the order of magnitude of the numerical and experimental cooling rates. To this end, we focus on the behaviors of the fictive temperature and $Q^n$ distributions, as these properties are traditionally challenging to be predicted properly by MD simulations.

Various functions have been proposed to describe the evolution of $T_f$ as a function of the cooling rate ($\gamma$). In particular, a Vogel-Fulcher temperature-dependence of the viscosity (or relaxation time) yields the following form[4]:

$$T_{\mathrm{f}} = T_0 - \frac{B}{\ln{(\gamma A)}} \quad \text{Eq. 4}$$

where $T_0$ is the fictive temperature at infinitely small cooing rate, and $A$ and $B$ fitting parameters. Note that an Arrhenius behavior is obtained in the limit of $T_0 = 0$, namely:

$$T_{\mathrm{f}} = - \frac{B}{\ln{(\gamma A)}} \quad \text{Eq. 5}$$

Alternatively, mode-coupling theory[45] suggests a power law dependence as[4]:

$$T_{\mathrm{f}} = T_0 + (A\gamma)^{1/\delta} \qquad \text{Eq. 6}$$

where A and $\delta$ are fitting parameters. Finally, optimization problems can typically be well described by a logarithmic law[4,46]:

$$T_{\mathrm{f}} = T_0 + A \log(\gamma) \qquad \text{Eq. 7}$$

Fig. 12 shows the evolution of the computed fictive temperature as a function of the cooling rate, as compared to the experimental value obtained at low cooling rate[47]. As expected and observed experimentally[6], $T_{\mathrm{f}}$ is found to decrease with decreasing cooling rate. The computed values of $T_{\mathrm{f}}$ are then fitted to the functions mentioned above and extrapolated to lower cooling rates for comparison with the experimental data. Overall, the computed values are reasonably well fitted by Eq. 4 and Eq. 6. In contrast, the fits offered assuming an Arrhenius or logarithmic dependence are less satisfactory. Similar results were observed in the case of glassy silica[4]. However, when extrapolated toward lower cooling rates for comparison with the experimental value of $T_{\mathrm{f}}$ (733 K[47]), we note that both the Vogel-Fulcher and mode-coupling theories significantly over-predict $T_{\mathrm{f}}$, the fits using Eq. 4 and Eq. 6 yielding $T_0$ = 1050 and 1330 K, respectively. This is in agreement with the fact that the Vogel-Fulcher equation typically breaks down at low temperature[48]. On the other hand, the Arrhenius form offers an excellent extrapolation toward the experimental data. Note that this type of Arrhenius dependence is typically observed experimentally[6].

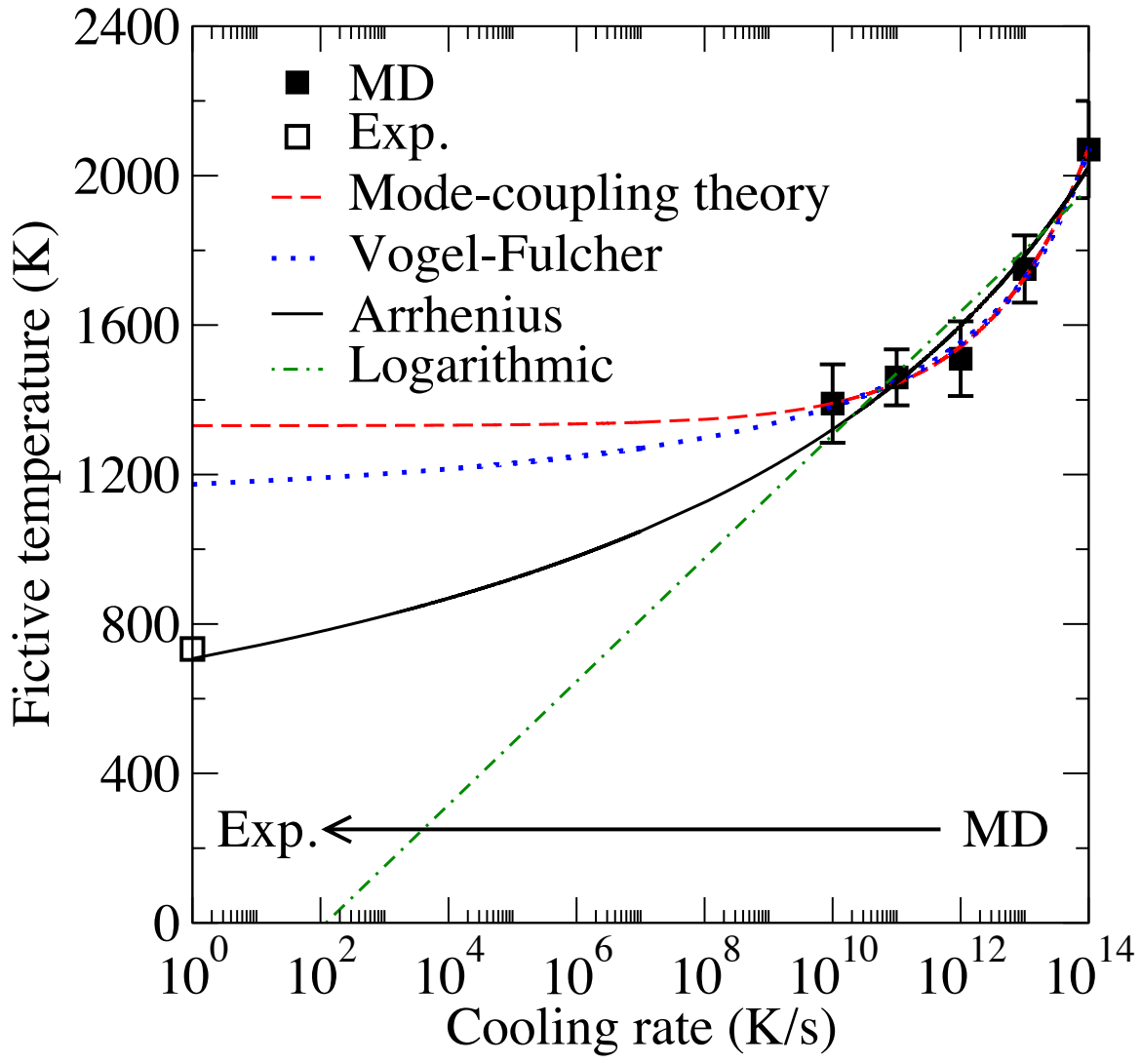

*Fig. 12: Experimental[47] and computed fictive temperature ($T_f$) as a function of the cooling rate ($\gamma$). The computed data are fitted following mode-coupling theory (Eq. 6, red dashed line), the Vogel-Fulcher equation (Eq. 4, blue dotted line), the Arrhenius equation (Eq. 5, black solid line), and a simple logarithmic law (Eq. 7, green dashed-dotted line). The obtained fits are extrapolated toward lower cooling rates for comparison with the experimental values[47].*

Coming back to the structure of the glass, the $Q^n$ distribution is arguably the structural feature of silicate glasses that is the most challenging to predict for MD simulations, which usually yield values that closely follow the predictions from a random model[19]. Note that the usage of very large systems did not offer any significant improvements in this matter[49]. Fig. 13 shows the fraction of each $Q^n$ species as a function of the cooling rate. We note that the computed data can be well fitted by a simple logarithmic function. As mentioned before, this type of dependence can be predicted by considering the cooling process as an optimization problem[4,46]. When extrapolated toward lower cooling rate, we observe that the logarithmic equation fits favorably with experimental values. A power law fit (not shown here) also yields a good agreement with the experimental data. Overall, these results suggest that, although MD simulations are intrinsically limited to high cooling rates due to the high computational cost, their outcomes can be directly compared to those obtained with more realistic cooling rates by extrapolation.

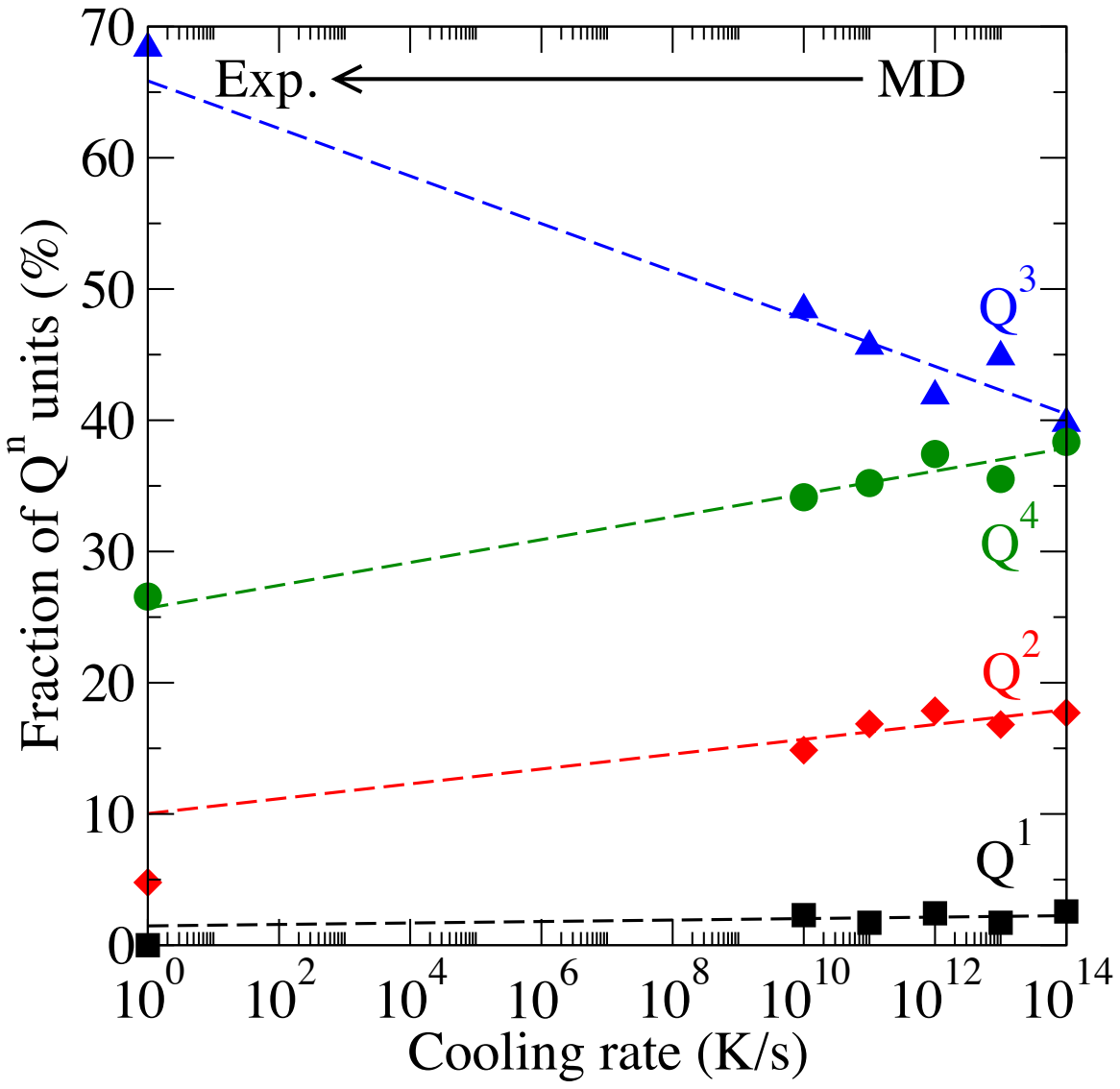

*Fig. 13: Experimental[43] and computed fractions of the $Q^n$ species in the glassy state as a function of the cooling rate. The computed fractions are fitted assuming a logarithmic law (Eq. 7) and extrapolated toward lower cooling rates for comparison with the experimental values*

## 2. Decoupling of enthalpy and volume relaxation

We now assess whether the evolutions of the molar enthalpy and volume for varying cooling rates are coupled to each other. To this end, we first fit the data by a power law, as suggested by mode-coupling theory[45]:

$$Y(\gamma) = Y_0 + (A\gamma)^{\delta} \qquad \text{Eq. 8}$$

where $Y(\gamma)$ refers to the molar enthalpy or volume and $Y_0$ its limit value for an infinitely small cooling rate. We then define a normalized "relaxation function" $f(\gamma)$ as:

$$f(\gamma) = \frac{Y(\gamma) - Y_0}{(A\gamma_{\max})^{\delta}} = \left(\frac{\gamma}{\gamma_{\max}}\right)^{\delta} \qquad \text{Eq. 9}$$

where $\gamma_{\max} = 10^{14}$ K/s, i.e., the maximum cooling rate considered herein. By definition, $f(0) = 0$ and $f(\gamma_{\max}) = 1$. The shape of the relaxation function is then captured by the exponent $\delta$, where a small value of $\delta$ indicates that the property slowly relaxes toward its limit value as the cooling rate decreases, whereas a large value of $\delta$ indicates that the property quickly relaxes toward its limit value.

Fig. 14 shows the relaxation function of the molar enthalpy and volume. Interestingly, we observe a clear decoupling between the relaxation of these two quantities, namely, the enthalpy ($\delta = 0.43$) converges toward its limit value much faster than the molar volume ($\delta = 0.12$). Such decoupling was also observed in the cases of the room-temperature relaxation of a silicate glass[34,50] and the response to irradiation of quartz[51,52]. Based on the results above, this decoupling can be understood as follows. On one hand, the enthalpy mostly depends on the short-range order of each atom (bond length, coordination number, angles) as the interaction energy between atoms and their second coordination shell is significantly lower than with their first coordination shell. Based on the results above, we observe that the short-range order structure of sodium silicate quickly plateaus as the cooling rate decreases. This arises from the fact that short-range order reorganizations do not involve any collective motion of atoms and, as such, are associated with low relaxation times. In turn, short-range defects are associated with a significant energy penalty. Altogether, the relaxation of the short-range order (and hence, of the enthalpy) is favored by kinetics and thermodynamics. On the other hand, the volume not only depends on the short-range order, but also on the medium-range order (FSDP, $Q^n$ distribution, rings, etc.). We observe that the medium-range order structural features of the glass show a greater dependence to the cooling rate. This arises from the fact that the relaxation of such features requires some collective reorganizations of atoms and, as such, is associated with longer relaxation times. In turn, since the medium-range order contribution to the energy of the system is low, there is no significant driving force for such relaxation. Hence, the relaxation of the medium-range order (and hence, of the volume) is not thermodynamically favored and exhibits a slow kinetics.

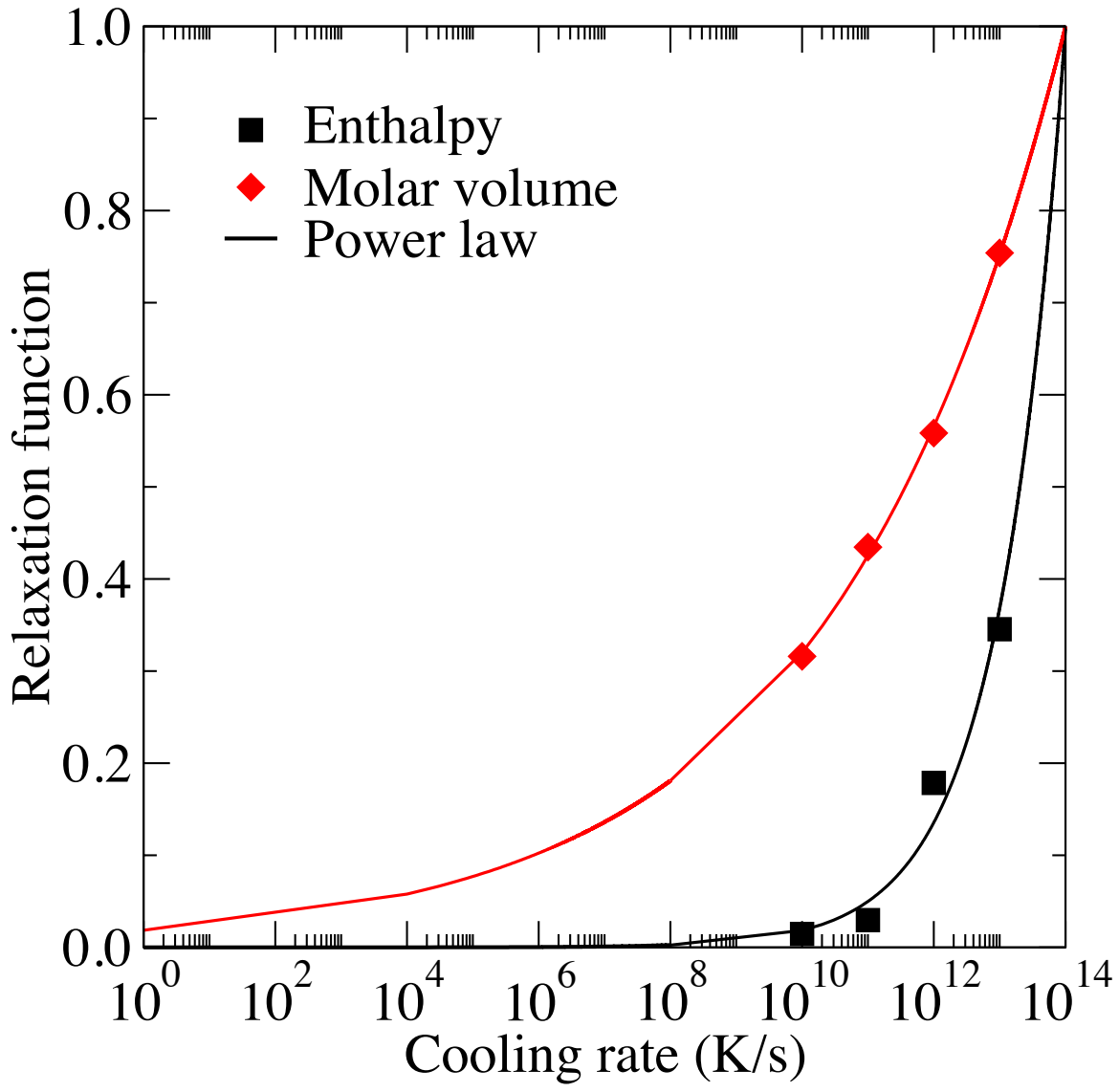

*Fig. 14: Relaxation function (see text) of the enthalpy and molar volume as a function of the cooling rate. The data are fitted by a power law (Eq. 8).*

## 3. Impact of the interatomic potential

Finally, these results allow us to establish some general conclusions regarding the impact of the interatomic potential (and of its parametrization method) on the comparison between glasses' simulations and experimental data. In general, the quality of the outcomes of MD simulations is intrinsically linked to that of the potential[53]. Besides their analytical forms, interatomic potentials can be classified in terms of the method used for their parametrization[54]. Namely, the interatomic potentials typically used to model glasses can be fitted to (i) experimental data (e.g., density, stiffness, etc.), (ii) first principle calculations (e.g., density functional theory or *ab initio* simulations), or (iii) a combination of both – note that the Teter potential used herein was parametrized based on the outcomes of density functional theory simulations.

Our results suggest that the outcomes of MD simulations relying on interatomic potentials fitted to first principle calculations can be successfully compared to experiments, when extrapolated toward lower cooling rates. In turn, this implies that simulated glasses (i.e., with unrealistically high cooling rates) should not be comparable to conventional glasses, but to hyperquenched glasses – although, unlike typical non-annealed hyperquenched glasses, MD-based glasses are free of any macroscopic stress if quenched within the *NPT* ensemble.

This raises the following question: are interatomic potentials fitted to experimental data obtained for glasses cooled using typical laboratory cooling rates realistic? Based on the present results, it appears that such an approach can offer reliable results if the potential is fitted to properties that weakly depend on the cooling rate. In contrast, relying on properties that are sensitive to the cooling rate (density, $Q^n$ distribution, etc.) to fit potentials is likely to yield unrealistic interatomic energies. In other terms, tuning potentials to "force" simulated glasses to be comparable to conventional experimental glasses instead of hyperquenched glasses will certainly offer an improved level of agreement, but this apparent agreement will likely arise from a "cancellation of errors."

An interesting example of this point is offered by the well-known van Beest–Kramer–van Santen (BKS) potential[55], which was fitted to a combination of *ab initio* and experimental data. In its original form, it offers a good prediction of α-quartz's structure, density, and stiffness[55]. However, the original parametrization of the BKS potential was found to yield an over-estimated density for glassy silica, when quenched at high cooling rate[4]. To solve this problem, it was proposed to truncate the short-range interactions by reducing the cutoff to 5.5 Å[4], which greatly improves the agreement between the simulated and experimental densities of glassy silica. However, this truncation results in the over-estimation of the Young's modulus of glassy silica[30], which indeed suggests that the fact the re-parametrization of the BKS potential yields a realistic density arises from a cancellation of errors. In contrast, it is interesting to note that recent simulations suggest that the original BKS potential (i.e., with an un-truncated cutoff of 10 Å) does indeed predict realistic density and Young's modulus for glassy silica, provided that a low cooling rate is used[8]. This demonstrates that forcing interatomic potentials to match glasses' experimental data might be counter-productive. A more accurate, yet more computationally depending approach would consist in systematically simulating glasses using a combination of several cooling rates and extrapolating the outcomes of the simulations towards realistic cooling rates to meaningfully compare them to experimental data.

## V. Conclusion

Overall, the cooling rate is found to affect the structure and properties of glassy sodium silicate. However, we find that the short-range order structure (bond lengths, coordination numbers, and angles) of the glass only weakly depends on the cooling rate, whereas the medium order appears to be more affected. This results in a decoupling between the relaxation of the enthalpy and volume with decreasing cooling rates. Further, despite the intrinsic limitation of MD simulations to the usage of high cooling rate, our simulations show that a handshake between experiments and MD simulations can be achieved by extrapolating the outcomes of the simulations toward lower cooling rates.

## Acknowledgements

This work was supported by the National Science Foundation under Grant No. 1562066 and by Corning Incorporated.

## References

[1] J. Du, in *Mol. Dyn. Simul. Disord. Mater.*, edited by C. Massobrio, J. Du, M. Bernasconi, and P.S. Salmon (Springer International Publishing, 2015), pp. 157–180.

[2] A. Pedone, G. Malavasi, A.N. Cormack, U. Segre, and M.C. Menziani, Theor. Chem. Acc. **120**, 557 (2008).

[3] X. Yuan and A.N. Cormack, J. Non-Cryst. Solids **283**, 69 (2001).

[4] K. Vollmayr, W. Kob, and K. Binder, Phys. Rev. B **54**, 15808 (1996).

[5] P.Y. Huang, S. Kurasch, J.S. Alden, A. Shekhawat, A.A. Alemi, P.L. McEuen, J.P. Sethna, U. Kaiser, and D.A. Muller, Science **342**, 224 (2013).

[6] A.K. Varshneya, *Fundamentals of Inorganic Glasses* (Academic Press Inc, 1993).

[7] K. Vollmayr, W. Kob, and K. Binder, J. Chem. Phys. (1998).

[8] J.M.D. Lane, Phys. Rev. E **92**, 12320 (2015).

[9] G. Agnello and A.N. Cormack, J. Non-Cryst. Solids **451**, 146 (2016).

[10] Y. Yu, B. Wang, M. Wang, G. Sant, and M. Bauchy, J. Non-Cryst. Solids **443**, 148 (2016).

[11] C.A. Angell, J. Non-Cryst. Solids **131**, 13 (1991).

[12] J.C. Mauro, Fontiers Mater. **1**, 20 (2014).

[13] J.C. Mauro and E.D. Zanotto, Int. J. Appl. Glass Sci. **5**, 313 (2014).

[14] J.C. Mauro, C.S. Philip, D.J. Vaughn, and M.S. Pambianchi, Int. J. Appl. Glass Sci. **5**, 2 (2014).

[15] J.C. Mauro, A.J. Ellison, and L.D. Pye, Int. J. Appl. Glass Sci. **4**, 64 (2013).

[16] M. Bauchy, B. Guillot, M. Micoulaut, and N. Sator, Chem. Geol. **346**, 47 (2013).

[17] J. Du and A. Cormack, J. Non-Cryst. Solids **349**, 66 (2004).

[18] A.N. Cormack, J. Du, and T.R. Zeitler, Phys Chem Chem Phys **4**, 3193 (2002).

[19] M. Bauchy, J. Chem. Phys. **137**, 44510 (2012).

[20] S. Plimpton, J. Comput. Phys. **117**, 1 (1995).

[21] A.C. Wright, J. Non-Cryst. Solids **159**, 264 (1993).

[22] J. Du and L.R. Corrales, Phys. Rev. B **72**, 92201 (2005).

[23] J. Du and L.R. Corrales, J. Non-Cryst. Solids **352**, 3255 (2006).

[24] M. Micoulaut and M. Bauchy, Phys. Status Solidi B **250**, 976 (2013).

[25] M. Bauchy and M. Micoulaut, Phys. Rev. B **83**, 184118 (2011).

[26] M. Bauchy and M. Micoulaut, Phys. Rev. Lett. **110**, 95501 (2013).

[27] M. Bauchy and M. Micoulaut, Nat. Commun. **6**, 6398 (2015).

[28] B. Mantisi, M. Bauchy, and M. Micoulaut, Phys. Rev. B **92**, 134201 (2015).

[29] M. Bauchy, B. Wang, M. Wang, Y. Yu, M.J. Abdolhosseini Qomi, M.M. Smedskjaer, C. Bichara, F.-J. Ulm, and R. Pellenq, Acta Mater. **121**, 234 (2016).

[30] B. Wang, Y. Yu, Y.J. Lee, and M. Bauchy, Front. Mater. **2**, 11 (2015).

[31] B. Wang, Y. Yu, M. Wang, J.C. Mauro, and M. Bauchy, Phys. Rev. B **93**, 64202 (2016).

[32] A. Pedone, G. Malavasi, A.N. Cormack, U. Segre, and M.C. Menziani, Chem. Mater. **19**, 3144 (2007).

[33] A.C. Wright, A.G. Clare, B. Bachra, R.N. Sinclair, A.C. Hannon, and B. Vessal, Trans Am Crystallogr Assoc **27**, 239 (1991).

[34] Y. Yu, M. Wang, D. Zhang, B. Wang, G. Sant, and M. Bauchy, Phys. Rev. Lett. **115**, 165901 (2015).

[35] B. Ruta, G. Baldi, Y. Chushkin, B. Rufflé, L. Cristofolini, A. Fontana, M. Zanatta, and F. Nazzani, Nat. Commun. **5**, (2014).

[36] P.G. Debenedetti and F.H. Stillinger, Nature **410**, 259 (2001).

[37] M. Wang, B. Wang, N.M.A. Krishnan, Y. Yu, M.M. Smedskjaer, J.C. Mauro, G. Sant, and M. Bauchy, J. Non-Cryst. Solids **455**, 70 (2017).

[38] H.F. Shermer, J. Res. Natl. Bur. Stand. **57**, (1956).

[39] G.K. White, J.A. Birch, and M.H. Manghnani, J. Non-Cryst. Solids **23**, 99 (1977).

[40] B. Vessal, G.N. Greaves, P.T. Marten, A.V. Chadwick, R. Mole, and S. Houde-Walter, Nature **356**, 504 (1992).

[41] M.M. Smedskjaer, M. Bauchy, J.C. Mauro, S.J. Rzoska, and M. Bockowski, J. Chem. Phys. **143**, 164505 (2015).

[42] M. Wang, B. Wang, T.K. Bechgaard, J.C. Mauro, S.J. Rzoska, M. Bockowski, M.M. Smedskjaer, and M. Bauchy, J. Non-Cryst. Solids **454**, 46 (2016).

[43] H. Maekawa, T. Maekawa, K. Kawamura, and T. Yokokawa, J. Non-Cryst. Solids **127**, 53 (1991).

[44] S.R. Elliott, J. Non-Cryst. Solids **182**, 40 (1995).

[45] W. Kob and H.C. Andersen, Phys. Rev. E **51**, 4626 (1995).

[46] D.A. Huse and D.S. Fisher, Phys. Rev. Lett. **57**, 2203 (1986).

[47] Y. Vaills, T. Qu, M. Micoulaut, F. Chaimbault, and P. Boolchand, J. Phys.-Condens. Matter **17**, 4889 (2005).

[48] J.C. Mauro, Y. Yue, A.J. Ellison, P.K. Gupta, and D.C. Allan, Proc. Natl. Acad. Sci. **106**, 19780 (2009).

[49] L. Adkins and A. Cormack, J. Non-Cryst. Solids **357**, 2538 (2011).

[50] Y. Yu, J.C. Mauro, and M. Bauchy, Am. Ceram. Soc. Bull. **96**, 34 (2017).

[51] B. Wang, N.M.A. Krishnan, Y. Yu, M. Wang, Y. Le Pape, G. Sant, and M. Bauchy, J. Non-Cryst. Solids **463**, 25 (2017).

[52] B. Wang, Y. Yu, I. Pignatelli, G. Sant, and M. Bauchy, J. Chem. Phys. **143**, 24505 (2015).

[53] C. Massobrio, J. Du, M. Bernasconi, and P.S. Salmon, editors , *Molecular Dynamics Simulations of Disordered Materials* (Springer International Publishing, Cham, 2015).

[54] L. Huang and J. Kieffer, in *Mol. Dyn. Simul. Disord. Mater.*, edited by C. Massobrio, J. Du, M. Bernasconi, and P.S. Salmon (Springer International Publishing, 2015), pp. 87–112.

[55] B.W.H. van Beest, G.J. Kramer, and R.A. van Santen, Phys. Rev. Lett. **64**, 1955 (1990).